\documentclass[twocolumn,showpacs,aps,prl,superscriptaddress]{revtex4}

\usepackage{graphicx}
\usepackage{dcolumn}
\usepackage{amsmath}
\usepackage{epsfig}
\usepackage[dvips]{feynmp}

\RequirePackage{xspace}

\usepackage{relsize}
\def\babar{\mbox{\slshape B\kern-0.1em{\smaller A}\kern-0.1em
    B\kern-0.1em{\smaller A\kern-0.2em R}}}

\def\tautau     {\ensuremath{\tau^+\tau^-}\xspace}

\def\ccbar {\ensuremath{c\overline c}\xspace}

\def\piz   {\ensuremath{\pi^0}\xspace}

\def\Kbar  {\kern 0.2em\overline{\kern -0.2em K}{}\xspace}

\def\Kz    {\ensuremath{K^0}\xspace}
\def\Kzb   {\ensuremath{\Kbar^0}\xspace}
\def\KzKzb {\ensuremath{\Kz \kern -0.16em \Kzb}\xspace}
\def\Kp    {\ensuremath{K^+}\xspace}
\def\Km    {\ensuremath{K^-}\xspace}

\def\KpKm  {\ensuremath{\Kp \kern -0.16em \Km}\xspace}
 
\def\KL    {\ensuremath{K^0_{\scriptscriptstyle L}}\xspace}

\def\Dbar    {\kern 0.2em\overline{\kern -0.2em D}{}\xspace}

\def\Dz      {\ensuremath{D^0}\xspace}
\def\Dzb     {\ensuremath{\Dbar^0}\xspace}
\def\DzDzb   {\ensuremath{\Dz {\kern -0.16em \Dzb}}\xspace}
\def\Dp      {\ensuremath{D^+}\xspace}
\def\Dm      {\ensuremath{D^-}\xspace}

\def\DpDm    {\ensuremath{\Dp {\kern -0.16em \Dm}}\xspace}
\def\Dstar   {\ensuremath{D^*}\xspace}

\def\Dstarm  {\ensuremath{D^{*-}}\xspace}

\def\B       {\ensuremath{B}\xspace}
\def\Bbar    {\kern 0.18em\overline{\kern -0.18em B}{}\xspace}

\def\BB      {\ensuremath{B\Bbar}\xspace} 
\def\Bz      {\ensuremath{B^0}\xspace}
\def\Bzb     {\ensuremath{\Bbar^0}\xspace}
\def\BzBzb   {\ensuremath{\Bz {\kern -0.16em \Bzb}}\xspace}
\def\Bu      {\ensuremath{B^+}\xspace}
\def\Bub     {\ensuremath{B^-}\xspace}

\def\Bpm     {\ensuremath{B^\pm}\xspace}

\def\BpBm    {\ensuremath{\Bu {\kern -0.16em \Bub}}\xspace}

\def\BorBbar    {\kern 0.18em\optbar{\kern -0.18em B}{}\xspace}
\def\DorDbar    {\kern 0.18em\optbar{\kern -0.18em D}{}\xspace}
\def\KorKbar    {\kern 0.18em\optbar{\kern -0.18em K}{}\xspace}

\mathchardef\Upsilon="7107
\def\Y#1S{\ensuremath{\Upsilon{(#1S)}}\xspace}

\mathchardef\Deltares="7101
\mathchardef\Xi="7104
\mathchardef\Lambda="7103
\mathchardef\Sigma="7106
\mathchardef\Omega="710A

\def\Deltabar{\kern 0.25em\overline{\kern -0.25em \Deltares}{}\xspace}
\def\Lbar{\kern 0.2em\overline{\kern -0.2em\Lambda\kern 0.05em}\kern-0.05em{}\xspace}
\def\Sigbar{\kern 0.2em\overline{\kern -0.2em \Sigma}{}\xspace}
\def\Xibar{\kern 0.2em\overline{\kern -0.2em \Xi}{}\xspace}
\def\Obar{\kern 0.2em\overline{\kern -0.2em \Omega}{}\xspace}
\def\Nbar{\kern 0.2em\overline{\kern -0.2em N}{}\xspace}
\def\Xb{\kern 0.2em\overline{\kern -0.2em X}{}\xspace}

\newcommand{\tev}{\ensuremath{\mathrm{\,Te\kern -0.1em V}}\xspace}
\newcommand{\gev}{\ensuremath{\mathrm{\,Ge\kern -0.1em V}}\xspace}
\newcommand{\mev}{\ensuremath{\mathrm{\,Me\kern -0.1em V}}\xspace}
\newcommand{\kev}{\ensuremath{\mathrm{\,ke\kern -0.1em V}}\xspace}
\newcommand{\ev}{\ensuremath{\mathrm{\,e\kern -0.1em V}}\xspace}
\newcommand{\gevc}{\ensuremath{{\mathrm{\,Ge\kern -0.1em V\!/}c}}\xspace}
\newcommand{\mevc}{\ensuremath{{\mathrm{\,Me\kern -0.1em V\!/}c}}\xspace}
\newcommand{\gevcc}{\ensuremath{{\mathrm{\,Ge\kern -0.1em V\!/}c^2}}\xspace}
\newcommand{\mevcc}{\ensuremath{{\mathrm{\,Me\kern -0.1em V\!/}c^2}}\xspace}

\def\invfb   {\ensuremath{\mbox{\,fb}^{-1}}\xspace}

\def\mus  {\ensuremath{\rm \,\mus}\xspace}

\def\mus        {\ensuremath{\,\mu{\rm s}}\xspace}    

\def\to                 {\ensuremath{\rightarrow}\xspace}

\def\pep2{PEP-II}

\def\gsim{{~\raise.15em\hbox{$>$}\kern-.85em
          \lower.35em\hbox{$\sim$}~}\xspace}
\def\lsim{{~\raise.15em\hbox{$<$}\kern-.85em
          \lower.35em\hbox{$\sim$}~}\xspace}

\xspace

\newcommand{\epjBase}        {Eur.\ Phys.\ Jour.\xspace}
\newcommand{\jprlBase}       {Phys.\ Rev.\ Lett.\xspace}
\newcommand{\jprBase}        {Phys.\ Rev.\xspace}
\newcommand{\jplBase}        {Phys.\ Lett.\xspace}

\newcommand{\nimBaseC}       {Nucl.\ Instr.\ and Methods\xspace}

\newcommand{\npBase}         {Nucl.\ Phys.\xspace}

\newcommand{\epj}       [1]  {\epjBase\ {\bf #1}}

\newcommand{\nim}       [1]  {\nimBaseC~{\bf #1}}

\newcommand{\np}        [1]  {\npBase\ {\bf #1}}

\newcommand{\jpl}       [1]  {\jplBase\ {\bf #1}}

\newcommand{\jprl}      [1]  {\jprlBase\ {\bf #1}}
\newcommand{\pr}        [1]  {\jprBase\ {\bf #1}}

\def\jetset74   {\mbox{\tt Jetset \hspace{-0.5em}7.\hspace{-0.2em}4}\xspace}

\newcommand{\BABARPubYear}    {04}
\newcommand{\BABARPubNumber}  {014}

\newcommand{\SLACPubNumber} {10458}

\def\figurebox#1#2#3{%
    \def\arg{#3}%
    \ifx\arg\empty
    {\hfill\vbox{\hsize#2\hrule\hbox to #2{\vrule\hfill\vbox to #1{\hsize#2\vfill}\vrule}\hrule}\hfill}%
    \else
    {\hfill\epsfbox{#3}\hfill}%
    \fi}

\begin{document}

\preprint{\babar-PUB-\BABARPubYear/\BABARPubNumber} 
\preprint{SLAC-PUB-\SLACPubNumber} 

\begin{flushleft}
\end{flushleft}

\title{
{\Large \bf \boldmath
Search for \Bz Decays to Invisible Final States and to $\nu\bar{\nu}\gamma$
}}

\author{B.~Aubert}
\author{R.~Barate}
\author{D.~Boutigny}
\author{F.~Couderc}
\author{J.-M.~Gaillard}
\author{A.~Hicheur}
\author{Y.~Karyotakis}
\author{J.~P.~Lees}
\author{V.~Tisserand}
\author{A.~Zghiche}
\affiliation{Laboratoire de Physique des Particules, F-74941 Annecy-le-Vieux, France }
% ...
\collaboration{The \babar\ Collaboration}
\noaffiliation
\date{\today}

\begin{abstract}
We establish upper limits on branching fractions for \Bz decays to final states where the decay products are purely invisible 
(\textit{i.e.}, no observable final state particles)
and for \Bz decays to $\nu\bar{\nu}\gamma$.  Within the Standard Model,
these decays have branching fractions
that are below current experimental sensitivity, but various models of physics 
beyond the Standard Model predict significant contributions from these channels.
Using 88.5 million \BB pairs collected at the \Y4S 
resonance by the \babar\ experiment at the \pep2
$e^{+}e^{-}$ storage ring at the Stanford Linear Accelerator Center, we 
establish upper limits at the 90\% confidence level of $22 \times 10^{-5}$ for the branching 
fraction of $\Bz \to$ invisible and $4.7 \times 10^{-5}$ for the branching fraction of $\Bz \to \nu\bar{\nu}\gamma$. 
\end{abstract}


\maketitle

This paper describes a novel search for ``disappearance decays'' of \Bz mesons~\cite{conjugates}, where the
\Bz decay contains no observable final state particles, or such ``invisible'' decay products plus a single photon.
Invisible decay products are particles that are neither charged nor would generate a signal in an electromagnetic calorimeter.
These include neutrinos, as well as exotic, hypothetical particles (such as neutralinos).
The rate for invisible $B$ decays is negligibly small within the Standard Model (SM) of particle physics, but 
can be larger in several models of new physics.  
The SM decay $\Bz \to 
\nu\bar{\nu}$, which would give such an invisible experimental signature, is strongly helicity-suppressed
by a factor of order $(m_{\nu}/m_{\Bz})^2$~\cite{BB}.  When combined with the 
weak coupling constant $G_F^2$,
the resulting branching fraction is necessarily 
well below 
the range of present experimental observability.  
The SM expectation for the $\Bz \to \nu\bar{\nu}\gamma$ branching fraction is 
predicted to be of order $10^{-9}$, with very little hadronic uncertainty~\cite{nunugam}. 
An experimental observation of an invisible $(+\;\gamma)$ decay of a \Bz
with current experimental sensitivity
would thus be a clear sign of beyond-SM physics, as it could not be accommodated within SM theoretical uncertainty.
No quantitative experimental bounds on \Bz to invisible or $\nu\bar{\nu}\gamma$ have been previously established;
however, a reinterpretation of
data used for previous constraints on $b \to s\nu\bar{\nu}$ and other modes 
could potentially
imply upper limits 
on the quark-level process of this decay~\cite{aleph}.

Several models of new physics can give significant branching fractions for invisible decays of the \Bz.  A phenomenological model motivated by 
the observation of an anomalous number of dimuon events 
by the NuTeV experiment
allows for an invisible \Bz decay to a $\bar{\nu}\chi^0_1$ final state, where $\chi^0_1$ is a neutralino, with a branching fraction in the $10^{-7}$ 
to $10^{-6}$ range~\cite{NuTeV,DDR}.  Also, models with large extra dimensions, which would provide a possible solution to the 
hierarchy problem, can 
also have the effect of producing significant, although small, rates for 
invisible \Bz decays~\cite{ADW,AW,DLP}.

The data used in this analysis were collected with the \babar\ detector at the \pep2 $e^{+}e^{-}$ collider.  
The data sample consists of an integrated luminosity of 
81.5 \invfb accumulated at the \Y4S resonance, containing $(88.5 \pm 1.0)$ million \BB pair events, 
and 9.6 \invfb accumulated at a center-of-mass (CM) energy about 30 \mev below \BB 
threshold. 
The asymmetric energies of the \pep2 $e^+$ and $e^-$ beams result in a Lorentz boost $\beta\gamma \approx 0.55$ of the \BB pairs.

A detailed description of the \babar\ detector is presented in Ref.~\cite{BABARNIM}.  Charged particle
momenta are measured in a tracking system consisting of a 5-layer double-sided silicon vertex tracker (SVT)
and a 40-layer hexagonal-cell wire drift chamber (DCH).  The
SVT and DCH operate within a 1.5 T solenoidal field, and have a combined solid angle coverage in the CM frame of 90.5\%.
Photons and long-lived neutral hadrons are detected and their energies are measured in a CsI(Tl) 
electromagnetic calorimeter (EMC), which
has a solid angle coverage in the CM frame of 90.9\%.  Muons 
are identified in the instrumented flux return (IFR), composed of resistive plate chambers and layers of iron that
return the magnetic flux of the solenoid.  A detector of internally reflected Cherenkov light (DIRC)
is used for identification of charged kaons and pions.  A {\tt GEANT4}~\cite{GEANT4} based Monte Carlo simulation 
of the \babar\ detector response was
used to optimize the signal selection criteria and evaluate the signal detection efficiency.

The detection of invisible $B$ decays uses the fact that 
$B$ mesons are created in pairs, due to flavor conservation in $e^{+}e^{-}$ interactions.
If one $B$ 
is reconstructed in an event, one can thus infer that another $B$
has been produced.  This technique has been exploited in several \babar\ 
analyses~\cite{KnunuConf,taunuSemilepConf,taunuHadrConf}.  
We reconstruct events in which a \Bz decays to $D^{(*)-}\ell^{+}\nu$,
then look for consistency with an invisible decay of the other neutral $B$
(no observable final state particles) in the rest of the event.  
In order
to help reject non-\BB background, R2, the ratio of the second and zeroth Fox-Wolfram moments, is required 
to be less than 0.5~\cite{fox}.

We reconstruct $D^{(*)-}$ in the decay modes $\Dm \to K^{+}\pi^{-}\pi^{-}$ and $\Dstarm \to \Dzb\pi^{-}$
where, in the latter case, \Dzb is reconstructed in the decay modes $K^{+}\pi^{-}$, $K^{+}\pi^{-}\piz$, or $K^{+}\pi^{-}\pi^{+}\pi^{-}$.
To form $D^{(*)-}$ candidates in these decay modes,
$K^{+}$ candidates are combined with other tracks and \piz candidates in the event.
We identify 
$K^{+}$ candidates using 
Cherenkov-light information from the DIRC and energy-loss information (d$E$/d$x$) from the DCH and SVT.
The \piz candidates are composed of pairs of photons in the EMC.  
Each photon must have a reconstructed energy above 30 MeV in the laboratory frame,
and the sum of their energies must be greater than 200 MeV.
The \piz candidates must have an invariant mass between 115 and 150 MeV/$c^2$.  A mass-constrained fit is
imposed on \piz candidates in order to improve the resolution on the reconstructed invariant mass of the parent $D$ meson.

We require the \Dzb and \Dm candidates to have reconstructed invariant masses within 20 \mevcc of their respective nominal 
masses~\cite{PDG2002},
except for \Dzb decays with a \piz daughter, which must be within 35 \mevcc of the nominal \Dzb mass. 
Mass-constrained fits are applied to \Dzb and \Dm candidates in order to improve the measurement of the
momentum of each $D$.  
The difference in reconstructed mass between \Dstarm decay candidates and their \Dzb daughters must be less than
150 \mevcc.
All $D^{(*)-}$ candidates must have a total
momentum between 0.5 and 2.5\gevc in the CM frame.

Tracks selected as lepton candidates must pass either electron or muon selection criteria.
We identify electron candidates using energy and cluster shape information from the EMC, and Cherenkov angle information
from the DIRC.  Muon candidates are identified using information from the IFR and EMC.
Both electrons and muons must also have a momentum of at least 1\gevc, and a minimum of 20 DCH measurements.

To select $\Bz \to D^{(*)-}\ell^{+}\nu$ candidates, we require a $D^{(*)-}$ candidate and a lepton candidate 
to be consistent with production at a common
point in space.  We then calculate the cosine of the angle between the $D^{(*)-}\ell^{+}$ and the hypothesized \Bz candidate,
under the assumption that the only particle missing is a neutrino:
\begin{equation}
\hspace*{-0.2cm}\cos \theta_{B,D^{(*)-}\ell^{+}} = \frac{(2\,E_{B} E_{{D^{(*)-}\ell^+}} -m^2_{{\B}} - m^2_{{D^{(*)-}\ell^+}})}
{2\,|\vec{p}_{B}| |\vec{p}_{{D^{(*)-}\ell^+}}|   }.
\end{equation}
The energy $E_{{D^{(*)-}\ell^+}}$ and mass $m_{{D^{(*)-}\ell^+}}$ of the $D^{(*)-}\ell^+$ combination are determined
from reconstructed momentum information, and $m_{B}$ is the nominal \Bz mass.
The \Bz momentum $|\vec{p}_{B}|$ and energy $E_{B}$
are determined from beam parameters. 
When the assumption that a neutrino is the only missing particle is incorrect, 
$\cos \theta_{B,D^{(*)-}\ell^{+}}$ can fall outside the region $[-1,1]$.  
We thus require the $D^{(*)-}\ell^{+}$ combination to satisfy $-2.5 < \cos \theta_{B,D^{(*)-}\ell^{+}} < 1.1$.
The asymmetric cut admits higher mass \Dstar states where the additional decay products are lost, 
and allows for detector energy and momentum resolution.
When more than one such $D^{(*)-}\ell^{+}$ candidate
is reconstructed in an event, the one with the smallest value of $|\cos \theta_{B,D^{(*)-}\ell^{+}}|$ is taken.
We reconstruct a total of 126108 $\Bz \to D^{(*)-}\ell^{+}\nu$ candidate 
events in the data sample, with a purity of approximately 66\%.

\begin{figure}[!t]
\begin{center}
\includegraphics[width=8.5cm]{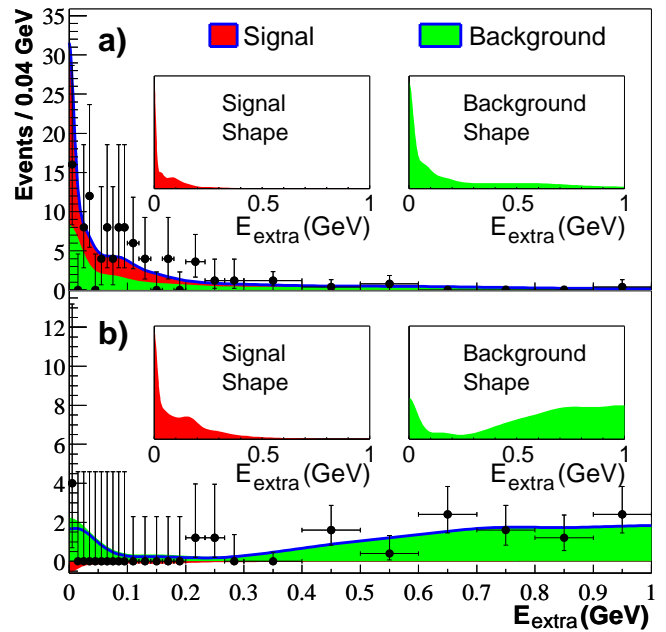}
\caption{
Distributions of $E_{\rm extra}$ for (a) \Bz $\to$ invisible and (b)
\Bz $\to \nu\bar{\nu}\gamma$.  The points with error bars correspond to data.
The curves represent maximum likelihood fits
to a sum of distributions modelling signal and background.  
}
\label{fig:likefits}
\end{center}
\end{figure}

We consider events with no charged tracks besides those of a $\Bz \to D^{(*)-}\ell^{+}\nu$ candidate.
Removing all decay products of the $D^{(*)-}\ell^{+}\nu$ candidate from consideration, we count the
number of remaining EMC clusters consistent with a \KL hypothesis, $N^{\rm extra}_{\KL}$,
and with a photon hypothesis, 
$N^{\rm extra}_{\gamma}$.
Due to 
accelerator-induced background and detector noise, the 
optimal requirements on $N^{\rm extra}_{\KL}$ and $N^{\rm extra}_{\gamma}$ are loose.
For \Bz $\to$ invisible candidates, we require that 
$N^{\rm extra}_{\KL} < 3$ and $N^{\rm extra}_{\gamma} < 3$.
For \Bz $\to \nu\bar{\nu}\gamma$ candidates, 
we require only that there be one remaining photon candidate
with energy greater than 1.2 \gev in the CM frame.

\begin{figure}[!t]
\begin{center}
\begin{minipage}[h]{4.47cm}
\includegraphics[width=4.47cm]{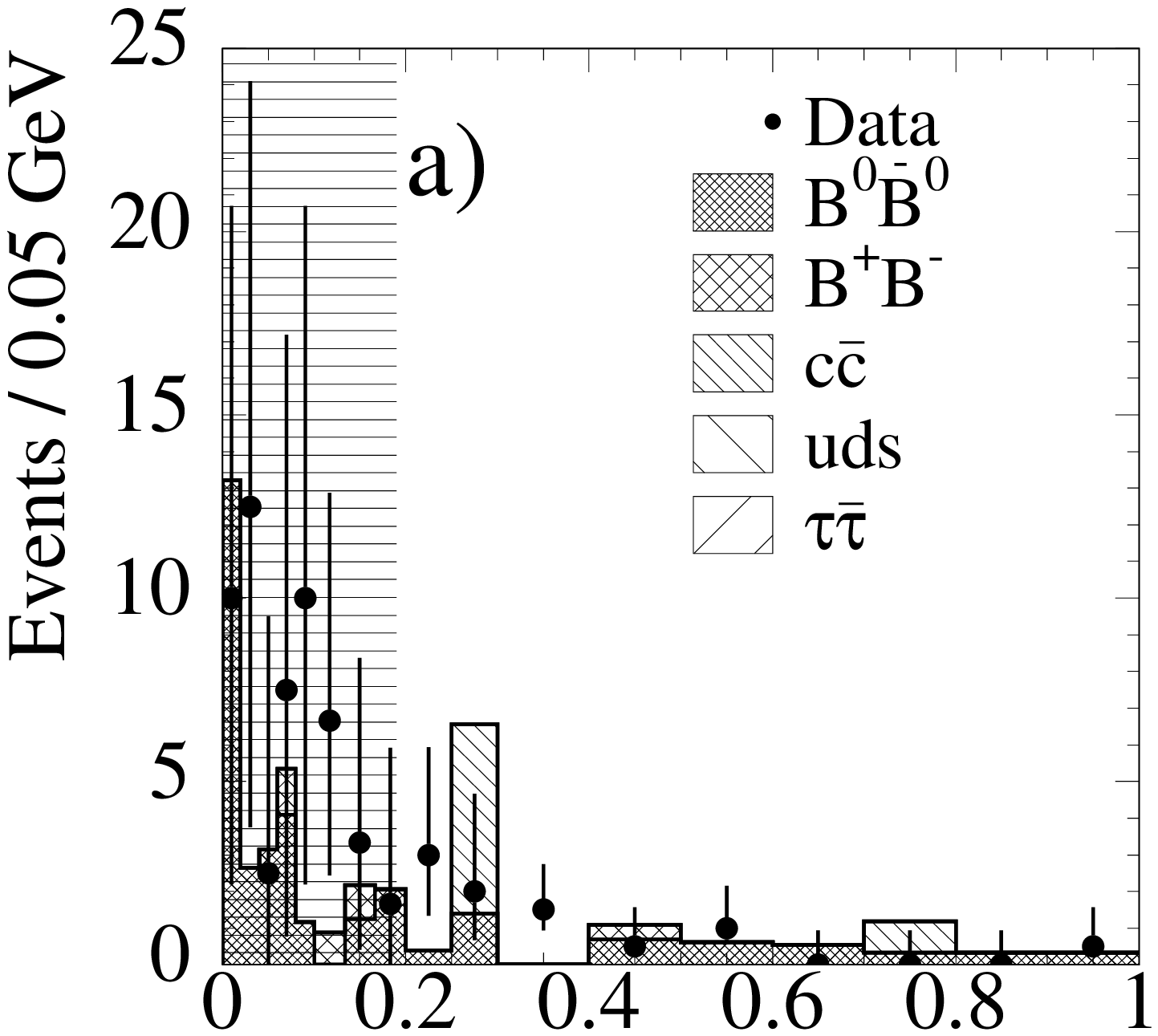}
\end{minipage}
\begin{minipage}[h]{4.03cm}
\vspace*{0.4cm}
\includegraphics[width=4.03cm]{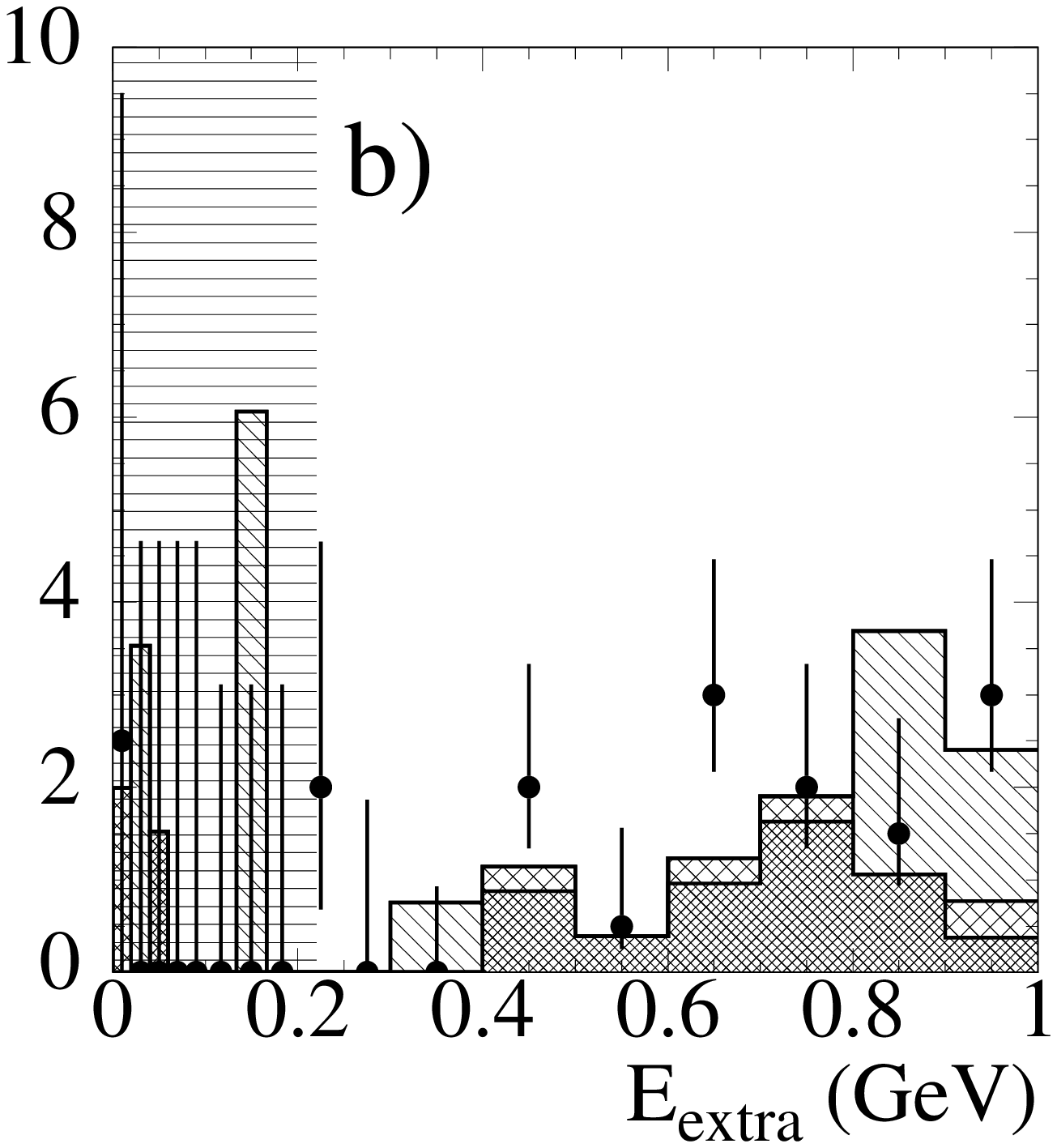}
\end{minipage}
\caption{
Comparison of $E_{\rm extra}$ between data (points with error bars) and Monte Carlo background simulation (histograms) for
(a) \Bz $\to$ invisible and (b) \Bz $\to \nu\bar{\nu}\gamma$.  The multiple categories of background
in the detector ($\Y4S \to \BzBzb$; $\Y4S \to \BpBm$; $e^{+}e^{-} \to \ccbar$; $e^{+}e^{-} \to u\bar{u}$, $d\bar{d}$, or $s\bar{s}$;
and $e^{+}e^{-} \to \tautau$ events) are each simulated by the Monte Carlo 
and plotted cumulatively.  No background from $e^{+}e^{-} \to u\bar{u}$, $d\bar{d}$, $s\bar{s}$, or \tautau is seen in the Monte Carlo 
sample.
For both of the modes, signal would tend to peak strongly in
the horizontally-shaded region.
}
\label{fig:etotleft}
\end{center}
\end{figure}

The total energy in the EMC, in the CM frame, of photon clusters that remain after the decay products of the 
$D^{(*)-}\ell^{+}\nu$ candidate are removed, is denoted by 
$E_{\rm extra}$.  For \Bz $\to \nu\bar{\nu}\gamma$, the energy of the highest-energy photon remaining in the event
(the hypothesized signal photon) is also removed from $E_{\rm extra}$.
In both \Bz $\to$ invisible and
\Bz $\to \nu\bar{\nu}\gamma$, this
variable is strongly peaked near zero for signal, whereas for the background it is less strongly peaked, 
as seen in Fig.~1.  
The background can peak near zero due to events in which all charged and
neutral particles from the signal \Bz are either outside the fiducial
volume of the detector, or are unreconstructed.
For \Bz $\to \nu\bar{\nu}\gamma$,
the background shape increases at large $E_{\rm extra}$ due to photons arising from
misreconstructed \piz decays, and
the best-fit amount of signal is slightly (but not significantly) negative.
We construct probability density functions (PDFs) for the $E_{\rm extra}$ distribution 
for signal ($\mathcal{F}_{\rm sig}$) and background ($\mathcal{F}_{\rm bkgd}$) using 
detailed simulation of signal and background data.  The background from accelerator and detector noise is 
modelled using randomly-triggered events in data.
The two PDFs are combined into an extended maximum likelihood function 
$\mathcal{L}$, defined as a function of the free parameters $N_{\rm sig}$ and $N_{\rm bkgd}$
\begin{eqnarray}
& & \mathcal{L}(N_{\rm sig}, N_{\rm bkgd}) = \frac{e^{-(N_{\rm sig} + N_{\rm bkgd})}}{N!} \; \times \nonumber\\
& & \qquad \quad \prod^{N}_{i=1} (N_{\rm sig}\mathcal{F}_{\rm sig}(E_i) + N_{\rm bkgd}\mathcal{F}_{\rm bkgd}(E_i)),
\end{eqnarray}
where $N_{\rm sig}$ and $N_{\rm bkgd}$ are the number of signal and background events, respectively.  The 
fixed parameters $N$ and $E_i$ are the total number of events in the data sample and 
the value of $E_{\rm extra}$ for the $i$th event, respectively.  The negative log-likelihood ($-\ln 
\mathcal{L}$) is then minimized with respect to $N_{\rm sig}$ and $N_{\rm bkgd}$ in the data sample.  
The resulting fitted values of $N_{\rm sig}$ and $N_{\rm bkgd}$ are 
$17\pm 9$ and $19^{+10}_{-8}$ 
for \Bz $\to$ invisible and 
$-1.1^{+2.4}_{-1.9}$ and $28^{+6}_{-5}$ 
for \Bz $\to \nu\bar{\nu}\gamma$, where the errors are 
statistical.
Figure~\ref{fig:etotleft} shows the $E_{\rm extra}$ distributions for 
\Bz $\to$ invisible and \Bz $\to \nu\bar{\nu}\gamma$.

Using detailed Monte Carlo simulation of \Bz $\to$ invisible and $\nu\bar{\nu}\gamma$ events,
we determine our signal efficiency to be $(16.7 \pm 1.0) \times 10^{-4}$
for \Bz $\to$ invisible and $(14.4 \pm 1.0) \times 10^{-4}$ for \Bz $\to \nu\bar{\nu}\gamma$, where the errors 
are again statistical.  For the \Bz $\to \nu\bar{\nu}\gamma$ channel, we assume a photon momentum 
distribution predicted by the constituent quark model for $\Bz \to \nu\bar{\nu}\gamma$ decay, as given in Ref.~\cite{nunugam}.
Of signal events that contain a reconstructed $\Bz \to D^{(*)-}\ell^{+}\nu$, approximately
46\% (30\%) of \Bz $\to$ invisible (\Bz $\to \nu\bar{\nu}\gamma$) events pass the signal selection.

We consider systematic uncertainties on the signal reconstruction efficiency, and also the uncertainty on the ratio of background to signal determined
in the fit.  Systematic uncertainties on the signal efficiency are dominated by the statistical size of the signal
Monte Carlo sample (resulting in relative uncertainties 
of 6.5\% and 6.8\% for \Bz $\to$ invisible and \Bz $\to \nu\bar{\nu}\gamma$, respectively) and by uncertainty on the efficiency
for determining the particle type of charged tracks (5.4\% for both channels).  Systematic uncertainty on the number of
signal events, due to uncertainty on the ratio of background to signal in the fit, is dominated by the parametrization
of the background and signal shapes (resulting in uncertainties on the number of signal events of 6.1 and
0.5 events for \Bz $\to$ invisible and \Bz $\to \nu\bar{\nu}\gamma$, respectively) and by the energy resolution
for reconstructing neutral clusters in the EMC (3.2 and 3.4 events, respectively).  
Other systematic uncertainties include the efficiency for reconstructing the charged tracks in the $\Bz \to D^{(*)-}\ell^{+}\nu$ decay, the 
charged track momentum resolution, and the total number of \BB events in the data sample.
The total systematic uncertainties on the
efficiency are $10.9\%$ and $11.1\%$, and on the fitted number of signal events are 
7.4 and 4.3 events, for \Bz $\to$ invisible and \Bz $\to \nu\bar{\nu}\gamma$, respectively.

To determine 90\% confidence level (C.L.) upper limits on the branching fractions of \Bz $\to$ invisible and \Bz $\to \nu\bar{\nu}\gamma$, 
we generate 8000 Monte Carlo experiments, each parametrized by the fitted numbers of signal and background events, the efficiency, and the number
of \BB events in the data sample.
Errors are incorporated into the simulated experiments via a convolution of the systematic effects (treated as Gaussian 
distributions) and the statistical error (taken from the non-Gaussian likelihood function from the fit).

\begin{figure}[!t]
\begin{center}
\begin{minipage}[h]{4.47cm}
\includegraphics[width=4.47cm]{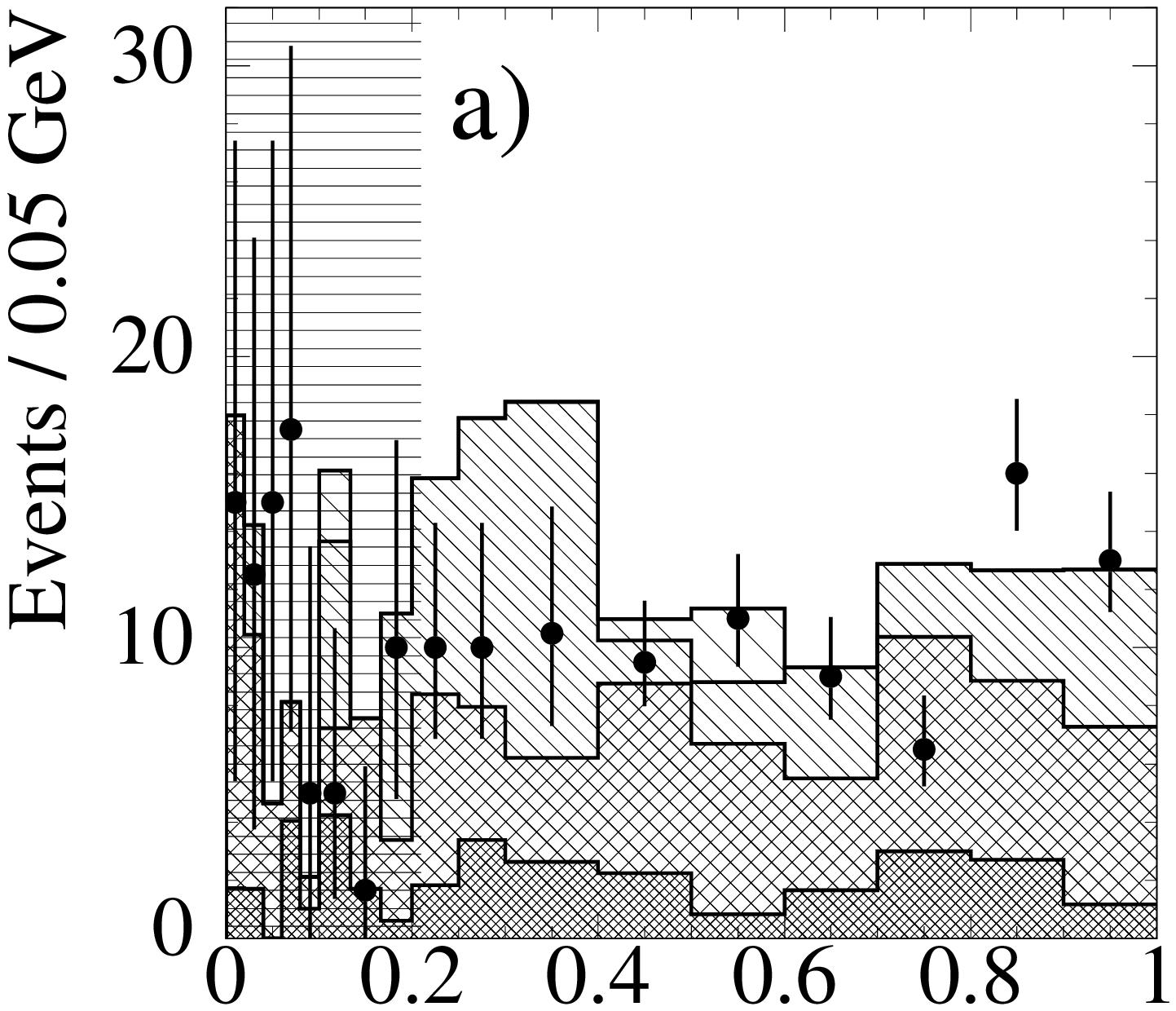}
\end{minipage}
\begin{minipage}[h]{4.03cm}
\vspace*{0.25cm}
\includegraphics[width=4.03cm]{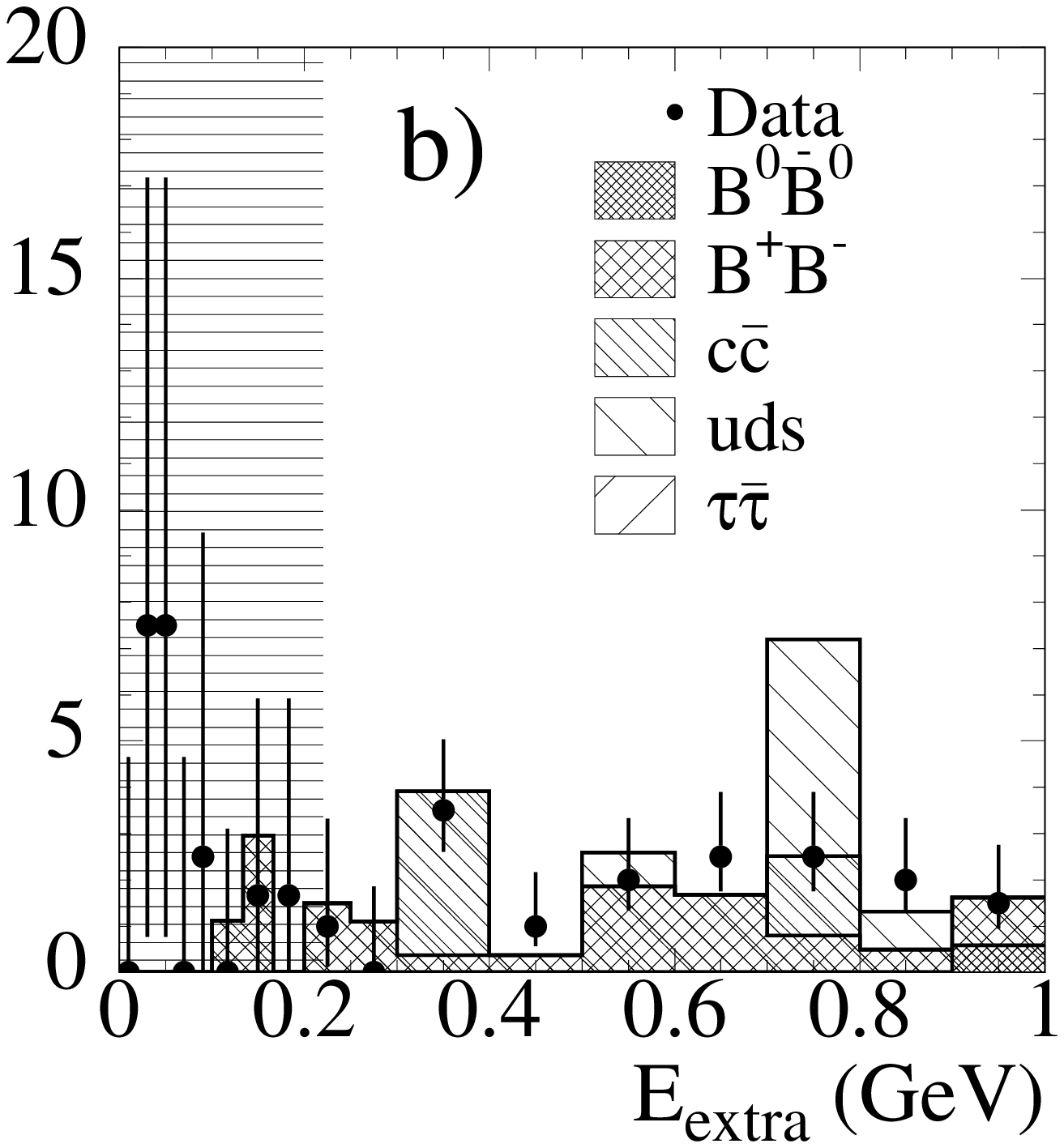}
\end{minipage}
\caption{
Comparison of $E_{\rm extra}$ between data (points with error bars) and Monte Carlo background simulation (histograms) for the validation 
channels
(a) \Bpm $\to$ invisible and (b) \Bpm $\to \nu\bar{\nu}\gamma$.  
No background from $e^{+}e^{-} \to \tautau$ is seen in the Monte Carlo sample.
}
\label{fig:bpmvalidation}
\end{center}
\end{figure}

The resulting upper limits on the branching fractions are
\begin{eqnarray}
\mathcal{B}(\Bz \to \mbox{invisible}) & < & 22 \times 10^{-5}\;\mbox{and}\nonumber\\
\mathcal{B}(\Bz \to \nu\bar{\nu}\gamma) & < & 4.7 \times 10^{-5}\;\mbox{at 90\% C.L.}\nonumber
\end{eqnarray}
If the \Bz $\to$ invisible branching fraction were zero, the probability of observing an 
equal or larger signal yield would be 6\%.

We perform 
validation cross-checks on the results of this analysis.  To check the
measurement of the efficiency for reconstructing $\Bz \to D^{(*)-}\ell^{+}\nu$ decays (which was determined using Monte Carlo simulation),
we select a data sample in which a \Bz and a \Bzb are both reconstructed as decays to $D^{(*)}\ell\nu$ in the same event.
Using the ratio of such ``double tag'' data events to events where just a single $D^{(*)}\ell\nu$ is reconstructed, and the number 
of \BzBzb events in the full data sample, we determine the efficiency for $\Bz \to D^{(*)-}\ell^{+}\nu$ reconstruction in data.  The result is 
consistent with that obtained from Monte Carlo simulation.

We also search for the unphysical modes \Bpm $\to$ invisible and \Bpm $\to \nu\bar{\nu}\gamma$ (which
would violate charge conservation), to check that their resulting signal is consistent with zero.  For these modes, we reconstruct 
\Bpm $\to D^0\ell\nu X^0$, where $X^0$ can be a photon, \piz, or nothing.  The \Dz is reconstructed in the same three decay modes
as in $\Bz \to D^{(*)-}\ell^{+}\nu$, and similar criteria are enforced for the reconstructed $B$ as for the neutral $B$ modes.  
All systematic errors are considered, and the ``double tags'' validation above is also performed for \Bpm reconstruction.  
The resulting fitted values of $N_{\rm sig}$ 
are $-6^{+10}_{-9}$(stat.)$\pm 6$(syst.)
for \Bpm $\to$ invisible
and $8^{+5}_{-4}$(stat.)$\pm 4$(syst.) 
for \Bpm $\to \nu\bar{\nu}\gamma$,
which are both consistent 
with zero.  
Figure~\ref{fig:bpmvalidation} shows the $E_{\rm extra}$ distributions for 
the two validation modes.

In summary, we obtain limits on branching fractions for \Bz decays to an invisible final state and for \Bz decays to 
$\nu\bar{\nu}\gamma$.
The upper limits 
at 90\% confidence level are
$22 \times 10^{-5}$ and $4.7 \times 10^{-5}$ for the \Bz $\to$ invisible and \Bz $\to \nu\bar{\nu}\gamma$
branching fractions, respectively.  
The latter limit assumes a photon momentum
distribution predicted by the constituent quark model for $\Bz \to \nu\bar{\nu}\gamma$ decay~\cite{nunugam},
whereas the \Bz $\to$ invisible limit is not decay-model dependent.

We are grateful for the excellent luminosity and machine conditions
provided by our \pep2\ colleagues, 
and for the substantial dedicated effort from
the computing organizations that support \babar.
The collaborating institutions wish to thank 
SLAC for its support and kind hospitality. 
This work is supported by
DOE
and NSF (USA),
NSERC (Canada),
IHEP (China),
CEA and
CNRS-IN2P3
(France),
BMBF and DFG
(Germany),
INFN (Italy),
FOM (The Netherlands),
NFR (Norway),
MIST (Russia), and
PPARC (United Kingdom). 
Individuals have received support from the 
A.~P.~Sloan Foundation, 
Research Corporation,
and Alexander von Humboldt Foundation.


\begin{thebibliography}{99}

\bibitem{conjugates}
Charge-conjugate decay modes are implied throughout this paper.

\bibitem{BB}
G.~Buchalla and A.J.~Buras, \np{B 400}, 225 (1993).

\bibitem{nunugam}
C.D.~Lu and D.X.~Zhang, 
\jpl{\bf B 381}, 348 (1996).

\bibitem{aleph}
ALEPH Collaboration, R.~Barate {\em et al.}, \epj{C 19}, 213 (2001).

\bibitem{NuTeV}
NuTeV Collaboration, T.~Adams {\em et al.}, 
\jprl{\bf 87}, 041801 (2001).

\bibitem{DDR}
A.~Dedes, H.~Dreiner, and P.~Richardson, 
\pr{\bf D 65}, 015001 (2002).

\bibitem{ADW}
K.~Agashe, N.G.~Deshpande, and G.-H.~Wu, 
\jpl{\bf B 489}, 367 (2000).

\bibitem{AW}
K.~Agashe and G.-H.~Wu, 
\jpl{\bf B 498}, 230 (2001).

\bibitem{DLP}
H.~Davoudiasl, P.~Langacker, and M.~Perelstein, 
\pr{\bf D 65}, 105015 (2002).

\bibitem{BABARNIM}
\babar\ Collaboration, B.\ Aubert {\em et al.}, \nim{A 479}, 1 (2002).

\bibitem{GEANT4}
{\tt GEANT4} Collaboration, S.~Agostinelli {\em et al.}, \nim{A 506}, 250 (2003).

\bibitem{KnunuConf}
\babar\ Collaboration, B.\ Aubert {\em et al.}, 
\babar-CONF-03/006,
hep-ex/0304020.

\bibitem{taunuSemilepConf}
\babar\ Collaboration, B.\ Aubert {\em et al.}, 
\babar-CONF-03/005,
hep-ex/0303034.

\bibitem{taunuHadrConf}
\babar\ Collaboration, B.\ Aubert {\em et al.}, 
\babar-CONF-03/004,
hep-ex/0304030.

\bibitem{fox}
G.C.~Fox and S.~Wolfram, \prl{\bf 41}, 1581 (1978).

\bibitem{PDG2002}
Particle Data Group, K.~Hagiwara {\em et al.}, \pr{\bf D 66}, 010001 (2002).

\end{thebibliography}
\end{document}